\documentclass[aip, amsmath, amssymb, reprint, floatfix]{revtex4-1}
\usepackage{graphicx}
\usepackage{dcolumn}
\usepackage{bm}
\usepackage[utf8]{inputenc}
\usepackage[T1]{fontenc}
\usepackage{mathptmx}
\usepackage{algorithm,algpseudocode}
\algnewcommand\Input{\item[\textbf{Input:}]}
\algnewcommand\Output{\item[\textbf{Output:}]}
\DeclareMathOperator{\E}{\mathbb{E}}
\DeclareMathOperator{\Var}{Var}

\begin{document}

\preprint{AIP/123-QED}

\title{One-to-one mapping between stimulus and neural state: \\Memory and classification}

\author{Sizhong Lan}\affiliation{$^1$China Mobile Research Institute, Beijing, 100053, China}\email{lsz@nzqrc.cn}

\date{\today}

\begin{abstract}
Synaptic strength can be seen as probability to propagate impulse, and according to synaptic plasticity, function could exist from propagation activity to synaptic strength.
If the function satisfies constraints such as continuity and monotonicity, the neural network under external stimulus will always go to fixed point, and there could be one-to-one mapping between the external stimulus and the synaptic strength at fixed point. In other words, neural network "memorizes" external stimulus in its synapses. A biological classifier is proposed to utilize this mapping.
\end{abstract}

\maketitle


\section{Introduction}

Known experiment results show that synaptic connection strengthens or weakens over time in response to increases or decreases in impulse propagation \cite{Plasticity1}.
It is also postulated that "neurons that fire together wire together" \cite{Hebb1,Hebb2}.
This biochemical mechanism, called synaptic plasticity \cite{Plasticity4,Plasticity5}, is believed to play a critical role in the memory formation \cite{mem_hypothesis4,mem_hypothesis1,mem_hypothesis2,mem_hypothesis3}, although it is still argued if synapse is the sole locus of learning and memory \cite{locus1,locus2}.
Meanwhile, a synapse propagates impulses stochastically \cite{Probabilistic4,Probabilistic2,Probabilistic1}, which means that synaptic strength could be measured with the probability of propagating an impulse successfully.
With this probabilistic treatment we find out that, in the plasticity process a synapse's strength would be inevitably attracted towards the same fixed point regardless of its initial strength, and for a neural network there could exist a one-to-one mapping between the external stimulus from environment and the synapses' strength at fixed point.
This one-to-one mapping serves the very purpose of ideal memory: to develop different stable neural state for different stimulus from the environment, and develop the same stable neural state for the same stimulus no matter what state is initialized with.
It follows that the synapses alone could sufficiently give rise to persistent memory: they could the sole locus of learning and memory.

The remainder of paper goes as follows.
Section \ref{section_connection} identifies the constraints under which synaptic plasticity of one synaptic connection leads to its fixed state and one-to-one stimulus-state mapping (memory).
Section \ref{section_nn} extends the concepts of fixed state and one-to-one mapping for the neural network consisting of many synaptic connections.
Section \ref{section_classifier} proposes a simple neural classifier utilizing this memory to classify handwritten digit images.

\section{Synaptic connection and its fixed point}
\label{section_connection}

\begin{figure}
\centering
\includegraphics[width=\linewidth]{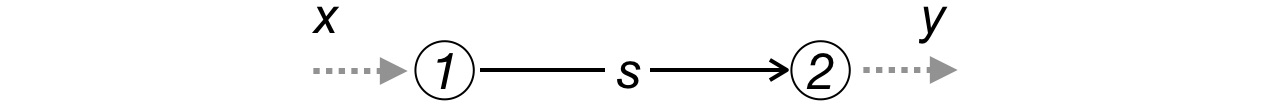}
\caption{A synaptic connection with strength $s$ is directed from neuron 1 to neuron 2.
The stimulus from environment or upstream neurons stimulates neuron 1 to fire action potential with probability $x$.
The synaptic connection propagates nerve impulse (action potential) to neuron 2.
As a result, neuron 2 is stimulated to fire with probability $y$.
That is, neuron 1 and 2 fire simultaneously ("fire together") with probability $y$.}
\label{fig:connection_model}
\end{figure}

Let us start with one synaptic connection as shown in FIG \ref{fig:connection_model}.
In nature, synapses are known to be plastic, low-precision and unreliable \cite{Probabilistic3}.
This stochasticity allows us to assume synaptic strength $s$ to be the probability (reliability) of propagating a nerve impulse through, instead of being weight (usually unbounded real number) as in Artificial Neural Network \cite{ANN1} (ANN).
Easily we have $y{=}xs$ where $x,s,y{\in}[0,1]$. Now we treat synaptic plasticity, i.e. the relation between synaptic strength $s$ and simultaneous firing probability $y$, as a function

\begin{equation}
s^*=\lambda(y).
\label{lambda_func}
\end{equation}

Here $s^*{\in}[0,1]$ represents the target value that a connection's strength will be strengthened or weakened to if the connection is under constant simultaneous firing probability $y$ (while $s$ in $y{=}xs$ represents current strength).
By $y{=}xs$ and Eq. (\ref{lambda_func}), we have $s^*{=}\lambda(xs)$ stating that, under constant stimulus probability $x$, the connection initialized with strength $s$ will evolve towards $s^*$.

Function $\lambda$ of Eq. (\ref{lambda_func}) truly links "firing together" and "wiring together".
For comparison, Hebbian learning rule \cite{Hebb1} treats synaptic plasticity, in the context of ANN, as a function ${\Delta}w{=}{\eta}xy$ to learn connections' weight from the training patterns; the function translates "firing together" into "neuron's input and output both being positive or negative".
Different from ANN, our model actually makes no assumption of neuron being computational unit, and aims to show that with $\lambda$ stimulus could sufficiently and precisely control the enduring fixed state of synaptic connection.
The following reasoning will hinge on this "target strength function" $\lambda$, and we will put constrains on this uncharted function to see how they affect the dynamics of connection strength and most importantly how stimulus is one-to-one mapped to the strength at fixated state.

\begin{figure}[!ht]
\centering
\includegraphics[width=\linewidth]{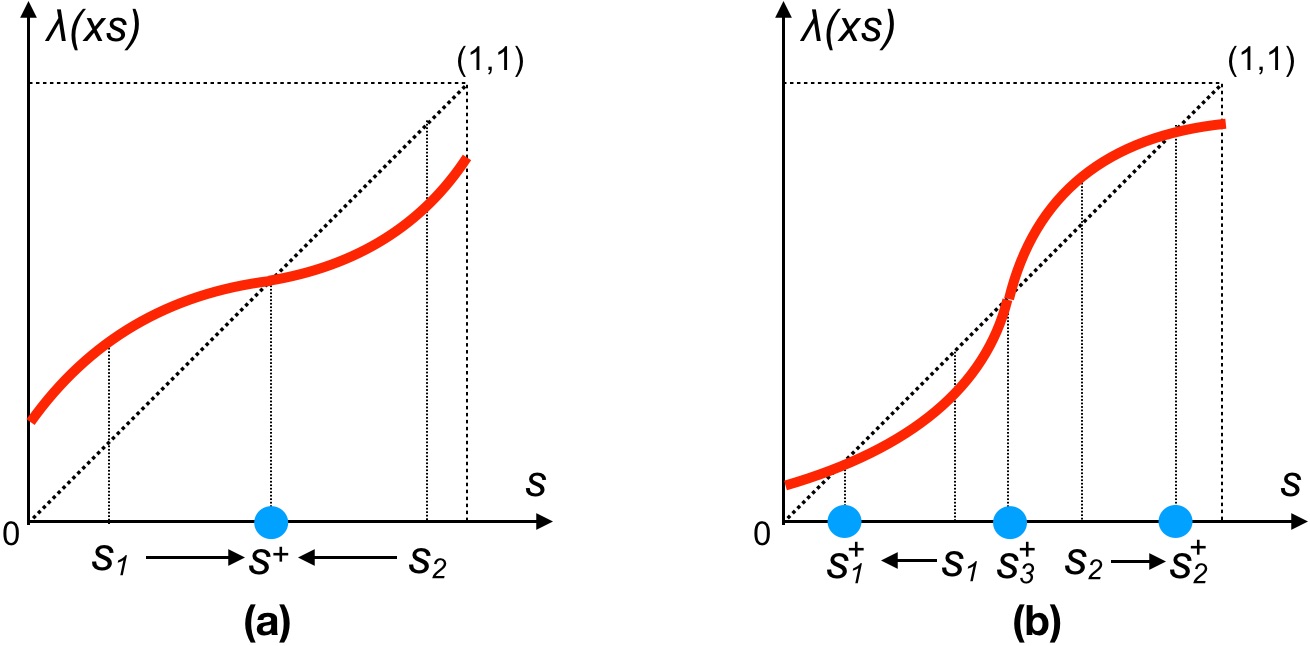}
\caption{Two examples of $\lambda(xs)$ are depicted as red bold lines, and their fixed points as blue dots.
\textbf{(a)} Given any initial $s_1{<}\lambda(xs_1)$, there must exist a fixed point $s^+{\in}(s_1,1]$; strength $s$ tends to increase from $s_1$ as long as target strength $\lambda(xs){>}s$.
Given any initial $s_2{>}\lambda(xs_2)$, there must exist a fixed point $s^+{\in}[0,s_2)$; strength $s$ tends to decrease from $s_2$ as long as target strength $\lambda(xs){<}s$. Controlled by these two tendencies, $s$ will reach and stay at fixed point $s^+$ such that $s^+{=}\lambda(xs^+)$.
\textbf{(b)} There are three fixed points $s_1^+$,$s_2^+$,$s_3^+$. Starting from any initial $s_1{\in}(s_1^+,s_3^+)$, strength decreases to $s_1^+$.
Starting from any initial $s_2{\in}(s_3^+,s_2^+)$, strength increases to $s_2^+$. Then strength tends to leave unstable fixed point $s_3^+$ for stable $s_1^+$ or $s_2^+$. Note that if countable fixed points exist for $\lambda(xs)$, one of them must be stable.}
\label{fig:tendency}
\end{figure}

\begin{algorithm}[H]
\begin{algorithmic}[1]
\Input{stimulus probability $x$, initial synaptic strength $s_0$, target strength function $\lambda$, strength step $\Delta_s$, and interations $I$.}
\Output{trajectory of strength $s$.}
\State initialize fire-together recorder ($10^4$-entries array): $recorder{\Leftarrow}0$.
\State initialize fire-together recorder pointer: $p{\leftarrow}0$.
\State initialize current strength: $s{\leftarrow}s_0$.
\State\textbf{for} $i{=}0$ \textbf{to} $I$ \textbf{do}
\State\hspace{0.5cm}preset current pointed entry of recorder: $recorder[p]{\leftarrow}0$.
\State\hspace{0.5cm}pick random $r1$ and $r2$ from uniform distribution $Unif(0, 1)$.
\State\hspace{0.5cm}\textbf{if} $x{>}r1$ \textbf{and} $s{>}r2$ \textbf{then}
\State\hspace{0.5cm}\hspace{0.5cm}neuron 1 and 2 fire together: $recorder[p]{\leftarrow}1$.
\State\hspace{0.5cm}\textbf{endif}
\State\hspace{0.5cm}\textbf{if} recorder has been traversed once ($i{\geq}10^4$) \textbf{then}
\State\hspace{0.5cm}\hspace{0.5cm}set $y$ with the proportion of $1$-entries in recorder.
\State\hspace{0.5cm}\hspace{0.5cm}set target strength: $s^*{\leftarrow}\lambda(y)$.
\State\hspace{0.5cm}\hspace{0.5cm}\textbf{if} $s^*{>}s$ \textbf{then}
\State\hspace{0.5cm}\hspace{0.5cm}\hspace{0.5cm}step-increase current strength: $s{\leftarrow}min(s{+}\Delta_s,1)$.
\State\hspace{0.5cm}\hspace{0.5cm}\textbf{end if}
\State\hspace{0.5cm}\hspace{0.5cm}\textbf{if} $s^*{<}s$ \textbf{then}
\State\hspace{0.5cm}\hspace{0.5cm}\hspace{0.5cm}step-decrease current strength: $s{\leftarrow}max(0,s{-}\Delta_s)$.
\State\hspace{0.5cm}\hspace{0.5cm}\textbf{end if}
\State\hspace{0.5cm}\textbf{end if}
\State\hspace{0.5cm}forward recorder pointer: $p{\leftarrow}(p{+}1){\bmod}10^4$.
\State\textbf{end for}
\end{algorithmic}
\caption{connection strength's tendency to fixed points}
\label{alg:connection_simulation_algorithm}
\end{algorithm}

\begin{figure}[!ht]
\centering
\includegraphics[width=\linewidth]{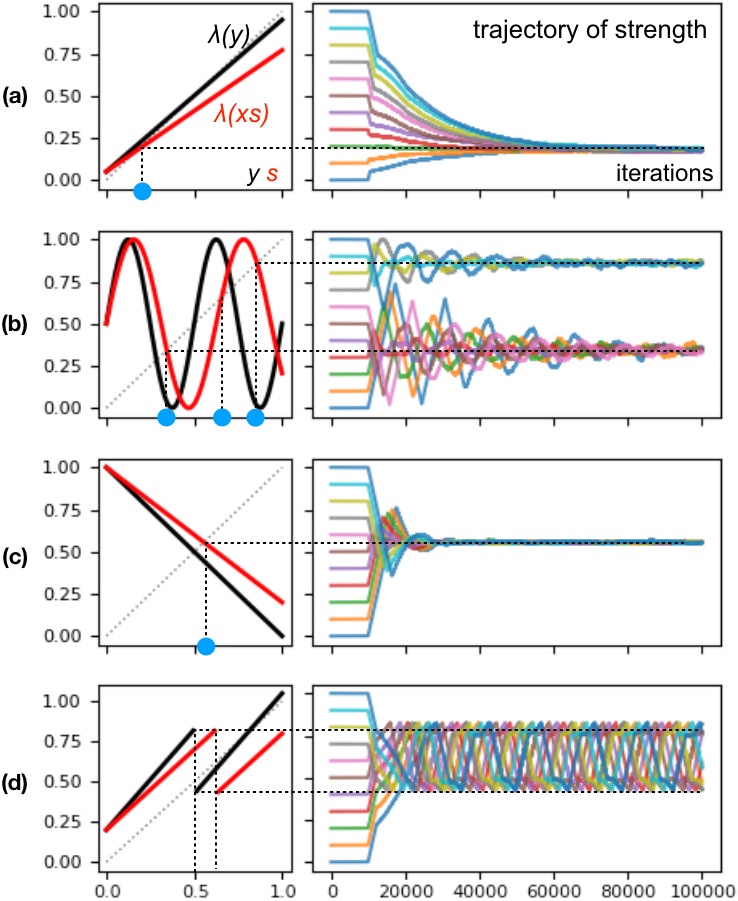}
\caption{The simulation results of Algorithm \ref{alg:connection_simulation_algorithm} for four typical $\lambda$ functions.
For each $\lambda$, eleven trails parameteried with incremental initial strength $s_0$ are run for $10^5$ iterations; all trails share the same stimulus probability $x{=}0.8$.
In each $\lambda$'s subfigure, the left chart depicts $\lambda(y)$ as black line, its horizontally scaled $\lambda(xs){=}\lambda(0.8s)$ in red line and fixed points as blue dots; the right chart shows the strength trajectories starting from incremental $s_0$.
\textbf{(a)} $\lambda(y){=}0.9y{+}0.05$. There exists one single fixed point for $\lambda(0.8s)$. All strength trajectories converge to this fixed point.
\textbf{(b)} $\lambda(y){=}0.5sin(4{\pi}y){+}0.5$. There are three fixed points for $\lambda(0.8s)$, two of which are stable ones for the trajectories to converge to with oscillation.
\textbf{(c)} $\lambda(y){=}{-}y{+}1$. All trajectories converge to one single fixed point.
\textbf{(d)} $\lambda(y)$ is discontinuous at $y{=}0.5$. There is no fixed point since there is no $s{\in}[0,1]$ such that $\lambda(0.8s){=}s$, and consequently the trajectories don't converge.
Instead, they fluctuate within "fixed interval", which as we will see is a useable compromise of fixed point. }
\label{fig:tendency_simulation}
\end{figure}

Here is our first constraint: \textbf{$\bm{\lambda}$ is continuous on $\bm{y}$}.
This constraint is neurobiologically justifiable regarding synaptic plasticity, since sufficiently small change in impulse probability would most probably result in arbitrarily small change in synaptic strength.
In that case, given any $x$, $\lambda(xs)$ is a continuous function on $s$ from unit interval $[0,1]$ to unit interval $[0,1]$, and according to Brouwer's fixed-point theorem \cite{Brouwer} there must exist a fixed point $s^+{\in}[0,1]$ such that $s^+{=}\lambda(xs^+)$: connection strength at $s^+$ will evolve to $s^+$ and hence fixate, no longer strengthened or weakened.
Here the crucial Brouwer's theorem is a fixed-point theorem in topology, which states that, for any continuous function $f(t)$ mapping a compact convex set (e.g. interval $[0,1]$ in our case; could be multi-dimensional) to itself, there is always a point $t^+$ such that $f(t^+){=}t^+$.
Moreover, as illustrated in FIG \ref{fig:tendency}, given any initial value the strength is always attracted towards fixed point.
Therefore, a gentle constraint of continuity on $\lambda$ function could preferably drive synaptic connection to the fixed state.

To verify connection strength's tendency towards fixed points, we design Algorithm \ref{alg:connection_simulation_algorithm} to simulate our connection model.
In this simulation\footnote{Source code can be found at \emph{https://github.com/lansiz/neuron}.}, recent simultaneous firings are recorded and the rate is supposed to approximate the simultaneous firing probability $y$; the connection updates its strength by a small step $\Delta_s{=}10^{-4}$ each iteration to the direction of target strength.
As shown in FIG \ref{fig:tendency_simulation}, we run the simulation for four typical target strength functions, and the strength trajectories resulted show that the constraint of continuity ensures the tendency towards fixed points given any initial strength.

\begin{figure}[!ht]
\centering \includegraphics[width=\linewidth]{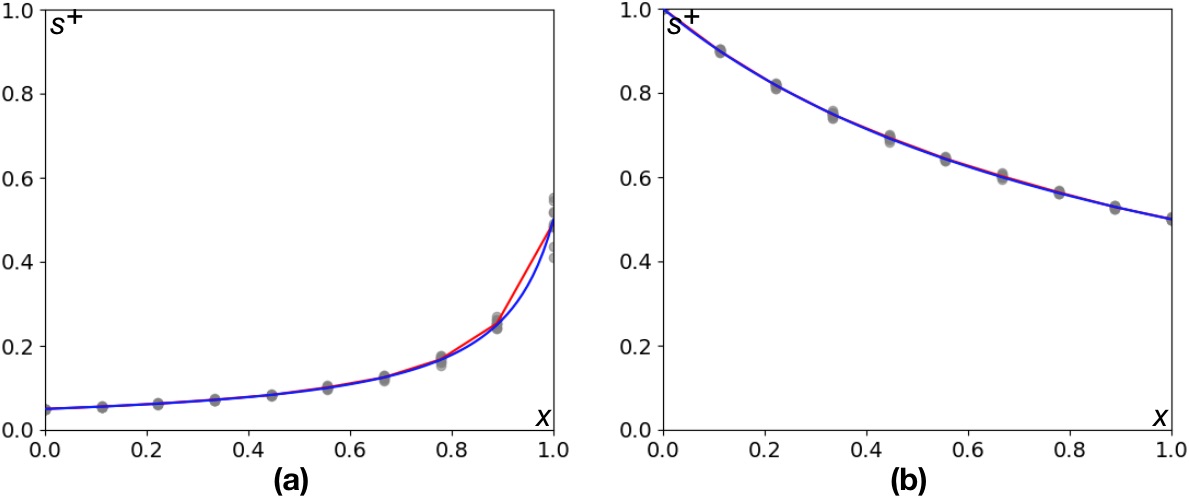} \caption{The simulation results of Algorithm \ref{alg:connection_simulation_algorithm} to reveal the relation of stimulus probability $x$ and fixed point strength $s^+$.
For each $\lambda$, simulation is parameterized with incremental $x$ (rather than $x{=}0.8$ as in FiG \ref{fig:tendency_simulation}) and randomized $s_0$.
Ten trails are run for each incremental $x$, and the ten converged $s$ values are averaged to be the $s^+$ value corresponding to its input $x$.
The red line represents the averaged $s^+$ values from simulation, while the blue line represents the true $s^+{=}\theta(x)$.
\textbf{(a)} Simulation is parameterized with $\lambda(y){=}0.9y{+}0.05$ and the results match $\theta(x){=}0.05{/}(1{-}0.9x)$ which is monotonically increasing.
\textbf{(b)} Simulation is parameterized with $\lambda(y){=}{-}y{+}1$ and the results match $\theta(x){=}1/(1{+}x)$ which is monotonically decreasing. }
\label{fig:theta_inverse}
\end{figure}

\begin{figure}[!ht]
\centering
\includegraphics[width=\linewidth]{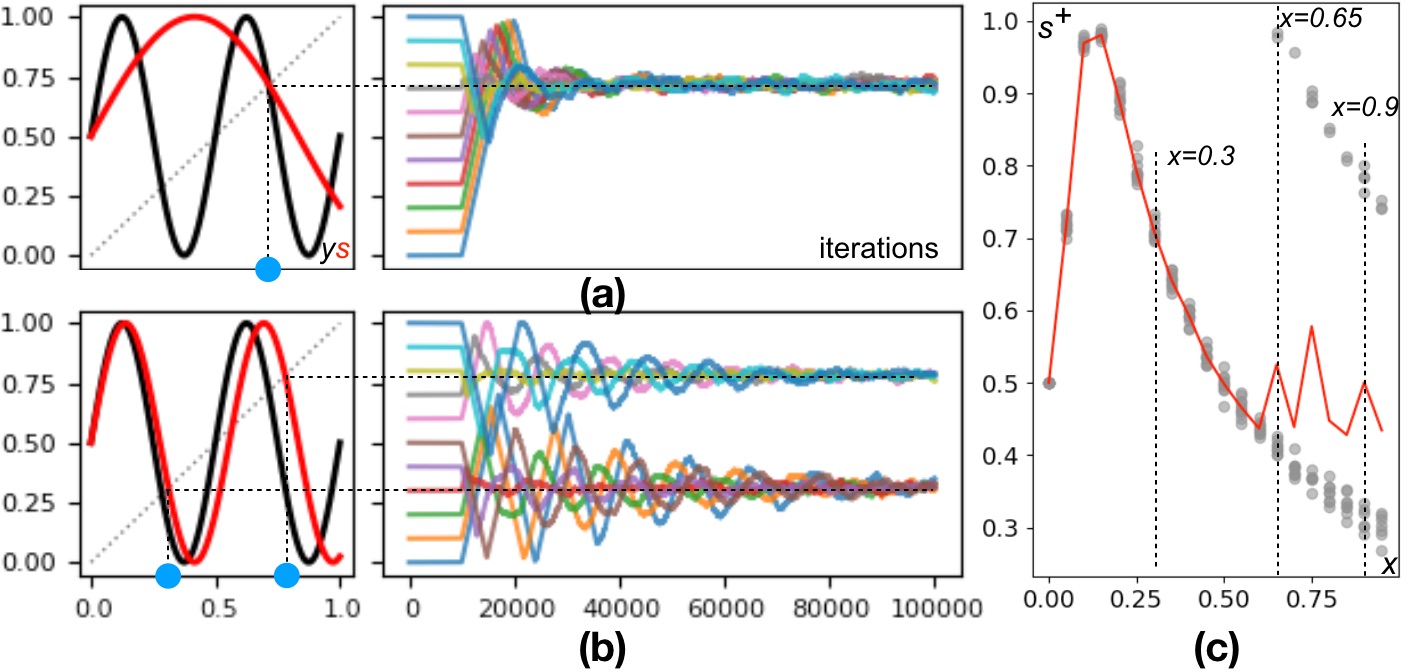}
\caption{The simulation results of Algorithm \ref{alg:connection_simulation_algorithm} for $\lambda(y){=}0.5sin(4{\pi}y){+}0.5$ in FIG \ref{fig:tendency_simulation}(b) to identify the relation between $x$ and $s^+$.
\textbf{(a)} Given $x{=}0.3$ there is one single fixed point regardless of initial strength $s_0$.
\textbf{(b)} As with $x{=}0.8$ in FIG \ref{fig:tendency_simulation}(b), given $x{=}0.9$ there are two stable fixed points.
Higher $s_0$ converges to higher fixed point; lower $s_0$ converges to lower one; No convergence to the middle unstable fixed point.
\textbf{(c)} Trails with incremental $x$ are run and the averaged $s^+$ values are depicted as in FIG \ref{fig:theta_inverse}. 
From $x{=}0.65$ and upwards, there are two possible stable fixed points to converge to depending on what value initial strength is randomized to, which means that there exists no $\theta$ function from $x$ to $s^+$. }
\label{fig:fp_different_x}
\end{figure}

\begin{figure}[!ht]
\centering
\includegraphics[width=\linewidth]{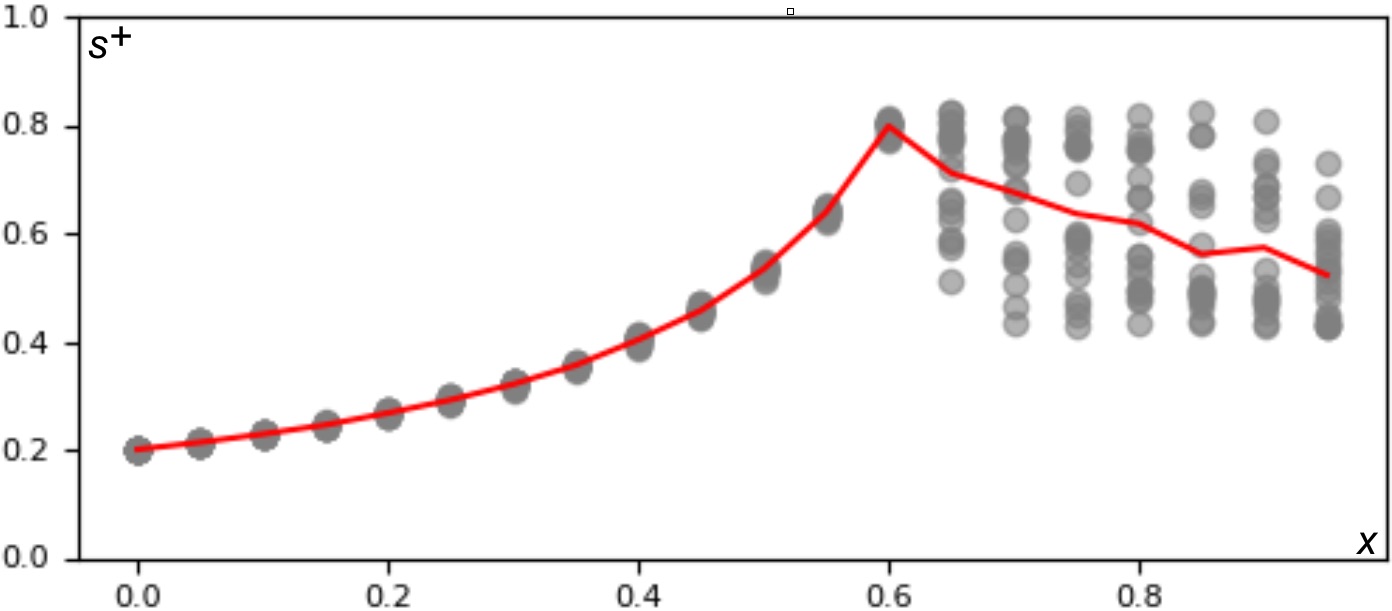}
\caption{The simulation results to find the $\theta$ function with respect to the discontinuous $\lambda$ in FIG \ref{fig:tendency_simulation}(d).
When $x{\gtrsim}0.6$, strength can evolve to any point within a "fixed interval" each time simulation is finished.
The absence of fixed point doesn't allow the existence of $\theta$. } \label{fig:theta_discont_pf}
\end{figure}

Our goal is to establish a one-to-one mapping between the stimulus and the connection strength at fixed point.
Specifically, we could \textbf{(1)} given any stimulus $x{\in}[0,1]$, identify the fixed point $s^+$ of connection strength without ambiguity; \textbf{(2)} given any strength $s^+{\in}[0,1]$ at fixed point, identify stimulus $x$ without ambiguity.
Among the four target strength functions in FIG \ref{fig:tendency_simulation}, $\lambda(y){=}0.9y{+}0.05$ and $\lambda(y){=}{-}y{+}1$ can lead to one-to-one stimulus-strength mapping.
Given any stimulus $x$, a synaptic connection equipped with one of these functions will have one single fixed point of strength regardless of its initial strength, such that the relation between stimulus and fixed point strength can be treated as a function $s^+{=}\theta(x)$.
In FIG \ref{fig:theta_inverse}, simulation shows that $\theta$ could be strictly monotonic and hence one-to-one mapping from $x$ to $s^+$, such that $\theta(x)$ has one-to-one inverse function $\theta^{-1}(s^+)$.
By contrast, FIG \ref{fig:fp_different_x} shows that $\lambda(y){=}0.5sin(4{\pi}y){+}0.5$ cannot ensure the uniqueness of fixed point and thus there is no such one-to-one $\theta(x)$; FIG \ref{fig:theta_discont_pf} shows that there is no $\theta$ either for the discontinuous $\lambda$ function in FIG \ref{fig:tendency_simulation}(d).

\begin{figure}[!ht]
\centering
\includegraphics[width=\linewidth]{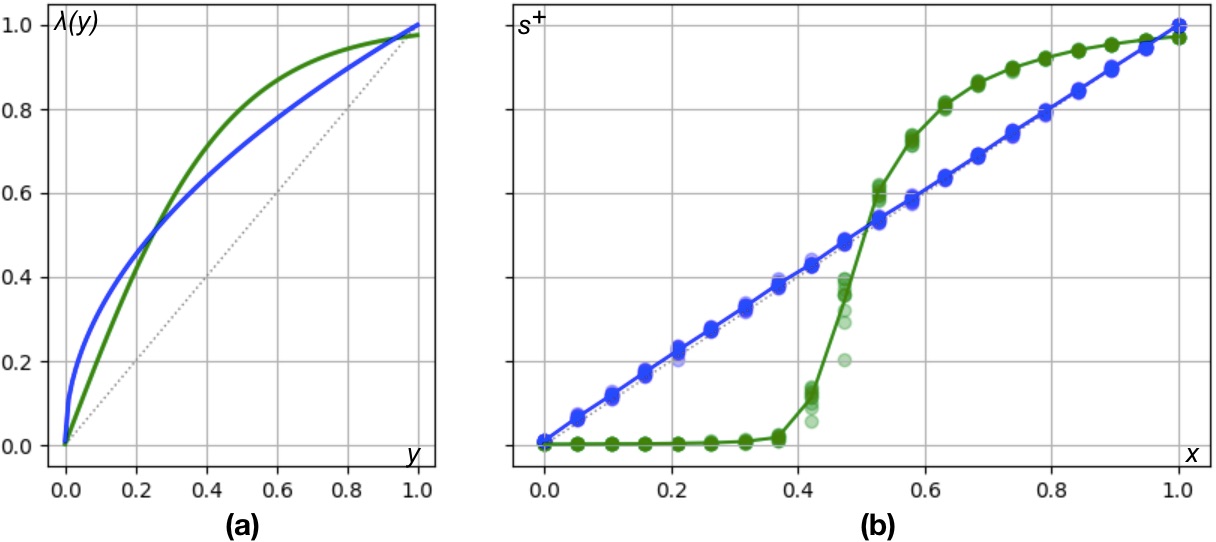}
\caption{\textbf{(a)} Our choices of $\lambda$ functions are $\lambda_L(y){=}0.99\sqrt{y}{+}0.01$ in blue and $\lambda_T(y){=}\frac{2}{1{+}e^{{-}4.4(y{+}0.01)}}{-}1$ in green.
Here $\lambda_T$ is a segment of shifted and scaled Sigmoid function.
They both obey the $\lambda$ constraints as discussed previously. \textbf{(b)} Simulation results show that, $\lambda_L$ leads to linear-like $\theta_L$ in blue such that $\theta_L(x){\approx}x$, and $\lambda_T$ leads to threshold-like $\theta_T$ in green. }
\label{fig:best_lambda}
\end{figure}

In fact, we can pinpoint more constraints on $\lambda$ as the conditions for function $\theta$ to be one-to-one mapping.
In addition to constraint of continuity, let \textbf{$\bm{\lambda(y)}$ be strictly monotonic on $\bm{[0,1]}$} and hence one-to-one; let \textbf{$\bm{\lambda(0){\neq}0}$} to rule out fixed point $s^+{=}0$.
In that case, $\lambda$ has inverse function $\lambda^{-1}(s)$ which is strictly monotonic between $\lambda(0)$ and $\lambda(1)$, and given any fixed point strength $s^+$ between we can identify stimulus $x{=}\lambda^{-1}(s^+){/}s^+$.
That is, function $\theta^{-1}(s^+)=\lambda^{-1}(s^+){/}s^+$ exists.
Let \textbf{$\bm{\lambda^{-1}(s){/}s}$ be strictly monotonic between $\bm{\lambda(0)}$ and $\bm{\lambda(1)}$}.
Then given any stimulus $x{\in}[0,1]$ there is one single fixed point $s^+$ such that $x{=}\lambda^{-1}(s^+){/}s^+$.
That is, function $s^+{=}\theta(x)$ exists.
Both of $\lambda(y){=}0.9y{+}0.05$ and $\lambda(y){=}{-}y{+}1$ obey all those constraints and their one-to-one $\theta$ functions can be verified by the simulation results in FIG \ref{fig:theta_inverse}, whereas $\lambda(y){=}0.5sin(4{\pi}y){+}0.5$ is not even strictly monotonic.
However, neither $\lambda(y){=}0.9y{+}0.05$ nor $\lambda(y){=}{-}y{+}1$ is ideal for our purpose.
Guided by these constraints, we choose $\lambda$ function carefully such that its derived $\theta(x)$ function is monotonically increasing and range of which spans nearly the entire $[0,1]$ interval, as shown in FIG \ref{fig:best_lambda}.
Of all the $\lambda$ constraints, continuity and strong monotonicity are reasonable requirements of consistency on the neurobiological process of synaptic plasticity, whereas $\lambda(0){\neq}0$ and strong monotonicity of $\lambda^{-1}(s){/}s$ are rather specific and peculiar claims.
Admittedly, those $\lambda$ constraints need to be supported by experimental evidences.

Now we have the one-to-one (continuous and strictly monotonic) functions $\lambda$, $\lambda^{-1}$, $\theta$ and $\theta^{-1}$, and in those functions initial strength $s_0$ is irrelevant.
Given $s^+$ we can identify $x$ and $y$ without ambiguity, and vice versa.
Our interpretation of these mappings is, the synaptic connection at fixed point precisely "memorizes" the information of what (stimulus) it senses and how it responses (with impulse propagation).

\section{Neural network and its fixed point}
\label{section_nn} 
\begin{figure}[!ht]
\centering
\includegraphics[width=\linewidth]{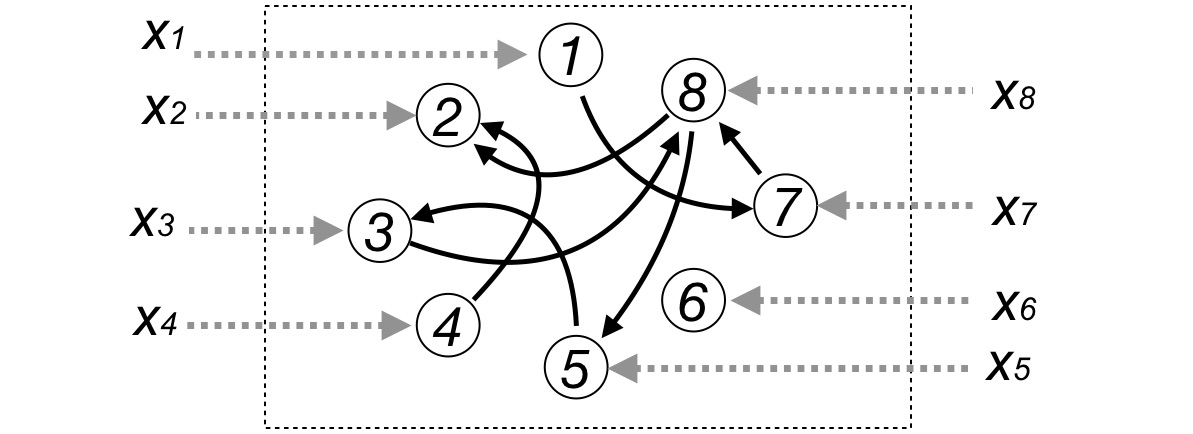}
\caption{A neural network (of one or multiple agents) consists of $n{\geq}2$ neurons and $c{\geq}1$ directed synaptic connections.
An example of $n{=}8$ and $c{=}7$ is depicted.
Each neuron receives stimulus from the environment with probability and propagates out nerve impulses throughout the synaptic connections, e.g., triggered by stimulus neuron $1$ propagates impulses stochastically down along the directed paths $1{\leadsto}7{\leadsto}8{\leadsto}2$ and $1{\leadsto}7{\leadsto}8{\leadsto}5{\leadsto}3$.
Cyclic path (e.g. $3{\leadsto}8{\leadsto}5{\leadsto}3$) is allowed and yet loop (e.g. $3{\leadsto}3$) isn't.
Each neuron could have either outbound or inbound connections, or neither, or both.}
\label{fig:neural_network_model}
\end{figure}

Now let us turn to the neural network shown in FIG \ref{fig:neural_network_model}.
A neural network could be treated as an "aggregate connection" as it turns out.
We shall see that, the definitions and reasoning for neural network align well with neural connection in last section.

As with synaptic connection, we can describe a neural network by defining 
\textbf{(1)} the external stimulus as an $n$-dimensional vector $X{\in}[0,1]^n$ in which each $x_i$ is the probability of neuron $i$ receiving stimulus; 
\textbf{(2)} the strength of all connections as a $c$-dimensional vector $S{\in}[0,1]^c$ in which each $s_{ij}$ is the strength of connection from neuron $i$ to neuron $j$ (denoted as $i{\leadsto}j$); 
\textbf{(3)} the simultaneous firing probabilities over all connections as a $c$-dimensional vector $Y{\in}[0,1]^c$ in which each $y_{ij}$ is the simultaneous firing probability over $i{\leadsto}j$.
In fact, one single neural connection is a special case of neural network with $c{=}1$ and $n{=}2$.

\begin{figure}[!ht]
\centering
\includegraphics[width=\linewidth]{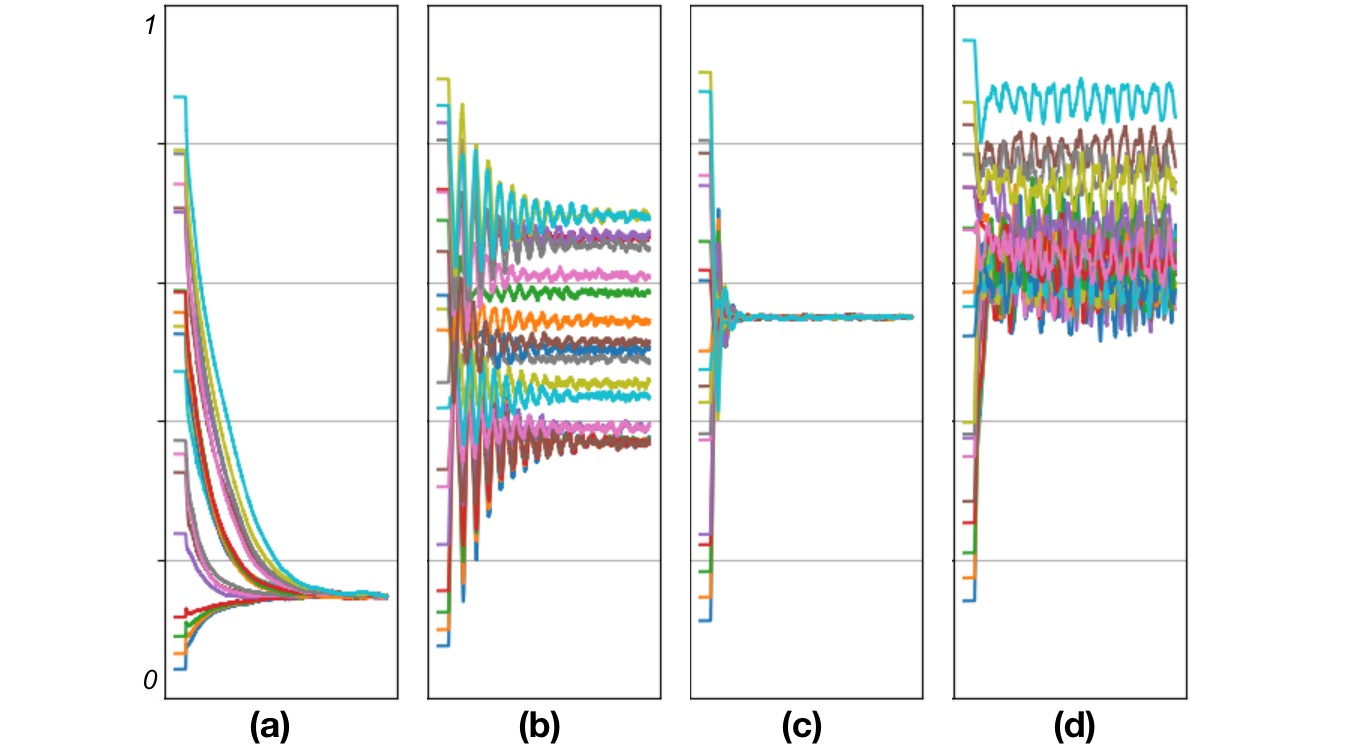}
\caption{Simulation results of neural network's tendency for the four typical $\lambda$ functions as in FIG \ref{fig:tendency_simulation}. The neural network has $n{=}8$ and $c{=}19$. The following observations hold true for any external stimulus and connections configuration:
\textbf{(a)} If all connections are equipped with $\lambda(y){=}0.9y{+}0.05$, the whole neural network has one single fixed point and the trajectories of mean of all connections' strength converge to one point.
\textbf{(b)} $\lambda(y){=}0.5sin(4{\pi}y){+}0.5$. Because each connection has two stable fixed points, there are $2^{19}$ stable fixed points for the whole neural network and $20$ possible convergence points of strength mean.
\textbf{(c)} $\lambda(y){=}{-}y{+}1$. There is one single fixed point for the neural network. The trajectories converge to one point.
\textbf{(d)} Discontinuous $\lambda$. The neural network has no fixed point as each synaptic connection has no fixed point. The trajectories don't converge. }
\label{fig:tendency_nn}
\end{figure}

Stimulus and strength uniquely determine impulses propagation within neural network, so there exists a mapping $\Psi{:}(X,S){\rightarrow}Y$.
Presumably, the mapping $\Psi$ is continuous on $S$. By Eq. (\ref{lambda_func}), there exists a mapping $\Lambda{:}Y{\rightarrow}S^*$ such that $s^*_{ij}{=}\lambda_{ij}(y_{ij})$ for each $y_{ij}$ in $Y$ and its counterpart $s^*_{ij}$ in $S^*$.
Here $S^*{\in}[0,1]^c$ is $c$-dimensional vector of connections' target strength, and mapping $\Lambda$ could be visualized as a vector of target strength functions such that entry $\Lambda_{ij}$ is $\lambda_{ij}$. Then with mapping $\Psi$ and $\Lambda$ we have a composite mapping $\Lambda{\circ}\Psi{:}(X,S){\rightarrow}S^*$.
If each $\lambda_{ij}$ function is continuous on its $y_{ij}$, mapping $\Lambda{\circ}\Psi$ must be continuous on $S$ and according to Brouwer's fixed-point theorem given $X$ there must exist one fixed point $S^+{\in}[0,1]^c$ such that $\Lambda{\circ}\Psi(X,S^+){=}S^+$.
And under constant stimulus $X$, neural network will go to fixed point $S^+$ as each connection $i{\leadsto}j$ goes to its fixed point $s^+_{ij}$.
Our simulation verifies that tendency as shown in FIG \ref{fig:tendency_nn}.
In this simulation, impulses traverse the neural network stochastically such that each neuron is fired at most once per iteration; synaptic connections update their strength as in Algorithm \ref{alg:connection_simulation_algorithm}.

Generally the number of stable fixed points for a neural network is $\prod^cf_{ij}$ where each $f_{ij}$ is the number of stable fixed points of $i{\leadsto}j$.
As in FIG \ref{fig:tendency_nn}(b), $\prod^cf_{ij}$ can be enormous when each $f_{ij}\geq2$.
As with synaptic connection, our goal is to establish one-to-one mapping between stimulus $X$ and fixed point $S^+$ for neural network and meanwhile keep initial strength $S_0$ out of picture.
$\lambda$'s continuity alone cannot ensure the uniqueness of fixed point, such that $S_0$ can determine which fixed point to go for.
Now with all the $\lambda$ constraints, we have:
\textbf{(1)} $\Lambda$ is a one-to-one mapping and thus has inverse mapping $\Lambda^{-1}{:}S^*{\rightarrow}Y$;
\textbf{(2)} there exists a mapping $\Theta{:}X{\rightarrow}S^+$, because under stimulus $X$ the neural network will go to the same unique fixed point $S^+$ no matter what initial strength $S_0$ to begin with;
\textbf{(3)} if $\Theta$ is a one-to-one mapping, $\Theta$ has inverse mapping $\Theta^{-1}{:}S^+{\rightarrow}X$.
With mapping $\Lambda$, $\Lambda^{-1}$, $\Theta$ and $\Theta^{-1}$ being one-to-one, given $S^+$ we can identify $X$ and $Y$ without ambiguity, and vice versa.
Therefore, the same interpretation with respect to synaptic connection could apply here: the neural network at fixed point precisely "memorizes" the information about the stimulus on many neurons and the impulse propagation across many connections.


Nevertheless, even all of $\lambda$ constraints are not sufficient to secure one-to-one $\Theta{:}X{\rightarrow}S^+$ for a neural network, as opposed to the neural connection. Here is a case. For $\Theta$ to be one-to-one, all neurons must have outbound connection.
Otherwise, e.g., for a neural network with three neurons (say $0$, $1$ and $2$) and two connections (say $0{\leadsto}1$ and $1{\leadsto}2$), stimulus $X_1{=}(1,1,0)$ and $X_2{=}(1,1,1)$ will result in the same fixed point because stimulus on neuron $3$, no matter what it is, affects no connection.
Or equivalently, for $\Theta$ to be one-to-one, the definition of $X$ should consider only the neurons with outbound connections such that $X$'s dimension $dim(X){\leq}n$.
In the perspective of information theory \cite{Shannon}, many-to-one $\Theta$ introduces equivocation to the neural network at fixed point, as if information loss occurred due to noisy channel.
If $dim(X){>}dim(S){=}c$, mapping $\Theta$ conducts "dimension reduction" on stimulus $X$, and information loss is bound to occur.

Here is a trivial case regarding stimulus dependence.
Consider a neural network with $0{\leadsto}2$, $1{\leadsto}2$ and $2{\leadsto}3$, and stimulus $X{=}(x_0,x_1)$.
When the neural network is at fixed point, $x_2{=}x_0s_{02}^+{+}x_1s_{12}^+{-}s_{02}^+s_{12}^+x_0\Pr(1|0)$ where $\Pr(1|0)$ is the probability of neuron $1$ being stimulated conditional on neuron $0$ being stimulated.
$\Pr(1|0){\neq}x_1$ if stimulus on neuron 1 and 2 are not independent.
$\Pr(1|0)$ affects $s_{23}^+$ and hence $S^+$, or in other words the neural network at fixed point gains the hidden information of $\Pr(1|0)$.
However, if $\Pr(1|0)$ varies, given mere $X$ there will be uncertainty about $S^+$ such that mapping $\Theta$ doesn't exist unless stimulus $X$ is "augmented" to $X{=}(x_0,x_1,\Pr(1|0))$.

\section{An application for classification}
\label{section_classifier}

Ideally, a neural network with memory of stimulus $X$ — formally, mapping $\Theta$ casts memory of stimulus $X$ as fixed point $S^+$ — should response to stimulus $X$ more "intensely" than the neural network with different memory responses to $X$.
Memory would manifest itself as impulses propagation throughout ensemble of neurons \cite{Hebb1,mem_manifest1,mem_manifest2,mem_manifest3}.
Thus, it is natural to differentiate response by counting the neurons fired or synaptic connections propagated by impulses.
Given the reasoning that synapse could be the sole locus of memory \cite{locus1,locus2}, we adopt the count of synaptic connections propagated as a macroscopic measure of how intensely memory responses to stimulus or stimulus "recalls" memory.
And accordingly we propose a classifier consisting of $g$ neural networks, which classifies stimulus into one of $g$ classes by the decision criteria of which neural network gets the most synaptic connections propagated.
Reminiscent of supervised learning \cite{Element}, each neural network of our classifier is trained to its fixed point by its particular training stimulus, and then a testing stimulus is tested on all $g$ neural networks independently to see which gets the most connections propagated.
For simplicity we assume testing itself doesn't jeopardize the fixed points of neural networks.
And most importantly we assume that for each neural network given any stimulus there is one single fixed point such that mapping $\Theta{:}X{\rightarrow}S^+$ exists.

Consider a neural network in the classifier to be trained by $\check{X}$ to fixed point $S^+$ and then tested by $X$.
In other words, neural network memorizing $\check{X}$ as $S^+$ is tested by $X$.
Because impulses propagate across the neural network stochastically, the count of synaptic connections propagated in one test should be random variable.
Let it be $Z_{\check{X}X}$.
Then for the neural network in FIG \ref{fig:neural_network_model} $Z_{\check{X}X}{=}\sum^{c}z_{ij}$ where each r.v. $z_{ij}{\sim}Bernoulli(x_is_{ij}^+)$, i.e., synaptic connection $i{\leadsto}j$ is propagated with probability $x_is_{ij}^+$ in the test such that $\Pr(z_{ij}{=}1){=}x_is_{ij}^+$.
Easily $z_{ij}$'s expected value is $\E[z_{ij}]{=}x_is_{ij}^+$, and its variance is $\Var(z_{ij}){=}x_is_{ij}^+(1{-}x_is_{ij}^+)$.
By central limit theorem, $Z$'s distribution could tend towards Gaussian-like (bell curve) as $c$ increases, even if all $z_{ij}$ are not independent and identically distributed. We have

\begin{equation}
\E[Z_{\check{X}X}]=\sum_{i{\leadsto}j}^{c}\E[z_{ij}]=\sum_{i{\leadsto}j}^{c}x_is_{ij}^+.
\label{Z_ev}
\end{equation}

And when $c$ is large,

\begin{equation}
\Var(Z_{\check{X}X})\approx\sum_{i{\leadsto}j}^{c}\Var(z_{ij})=\sum_{i{\leadsto}j}^{c}x_is_{ij}^+(1-x_is_{ij}^+).
\label{Z_var}
\end{equation}

For any $i{\leadsto}j$, in the training stage because $S^+{=}\Theta(\check{X})$ we have $s_{ij}^+{=}\theta_{ij}(\check{X})$, and in the testing stage $x_i$ is uniquely determined by $S^+$ and $X$ such that $x_i$ is a function of $\check{X}$ and $X$.

\begin{figure}[!ht]
\centering
\includegraphics[width=\linewidth]{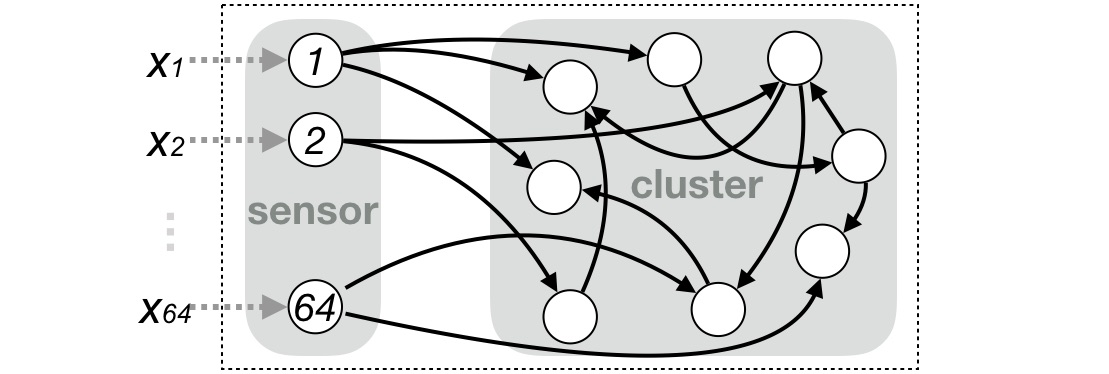}
\caption{The depicted neural network is basically the general one in FIG \ref{fig:neural_network_model} except that, to mimic real-life nervous system, an array of sensor neurons are specialized for receiving stimulus from no other neurons but the environment.
There are $64$ sensor neurons to accommodate $8{\times}8$-pixel image, and the rest are a cluster of $50$ neurons.
Each sensor neuron has $6$ outbound connections towards cluster, and each cluster neuron has $5$ outbound connections towards within cluster.
Connections are randomly put between neurons before training. }
\label{fig:brain1}
\end{figure}

\begin{figure}[!ht]
\centering
\includegraphics[width=\linewidth]{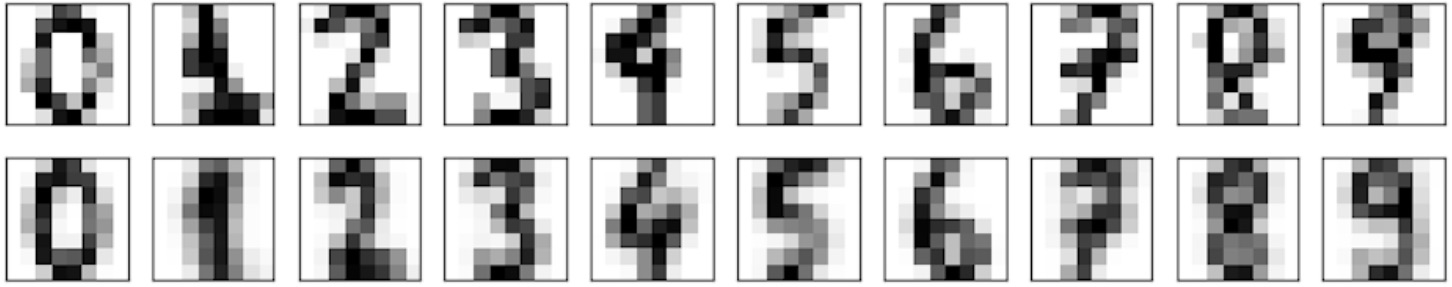}
\caption{A digit image has $8{\times}8{=}64$ pixels, and pixel grayscale is normalized to the value between $0$ and $1$ (by dividing $16$) as stimulus probability. The upper row shows samples of digit images, and the lower row shows the better written "average images", each of which is actually pixel-wise average of a set of images of a digit. Each neural network is trained in each iteration by the same "average image", or equivalently in each iteration by image randomly drawn from the set of images. }
\label{fig:average_images}
\end{figure}

\begin{figure}[!ht]
\centering
\includegraphics[width=\linewidth]{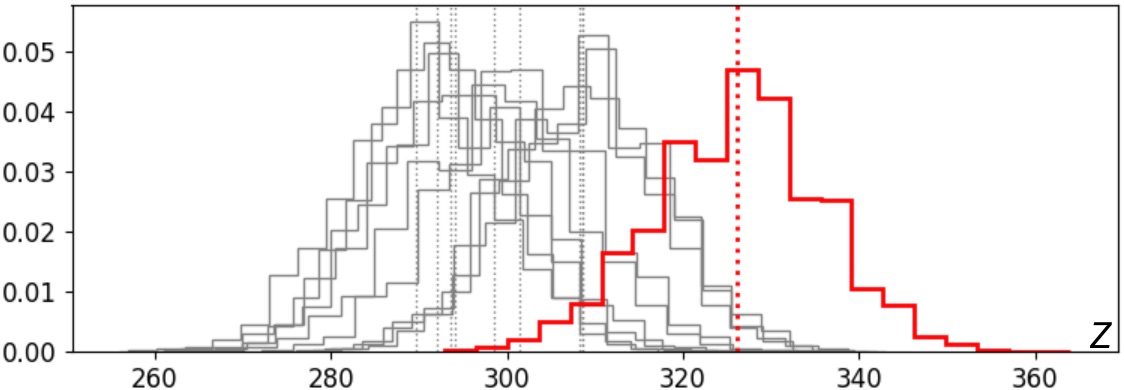}
\caption{The histogram (in probability density form) of $Z$. To collect $Z$ values, a digit-6 image is tested many times on each of the ten trained neural networks. All connections are equipped with $\lambda_T$. $Z_{66}$ of positive testing is in red, and the other nine $Z_{k6}$ of negative testing, where $k{=}0,1,2,3,4,5,7,8,9$, are in gray. $Z$'s sample mean for each digit is depicted as vertical dotted line. }
\label{fig:Z_dist_of_6}
\end{figure}

\begin{figure}[!ht]
\centering
\includegraphics[width=\linewidth]{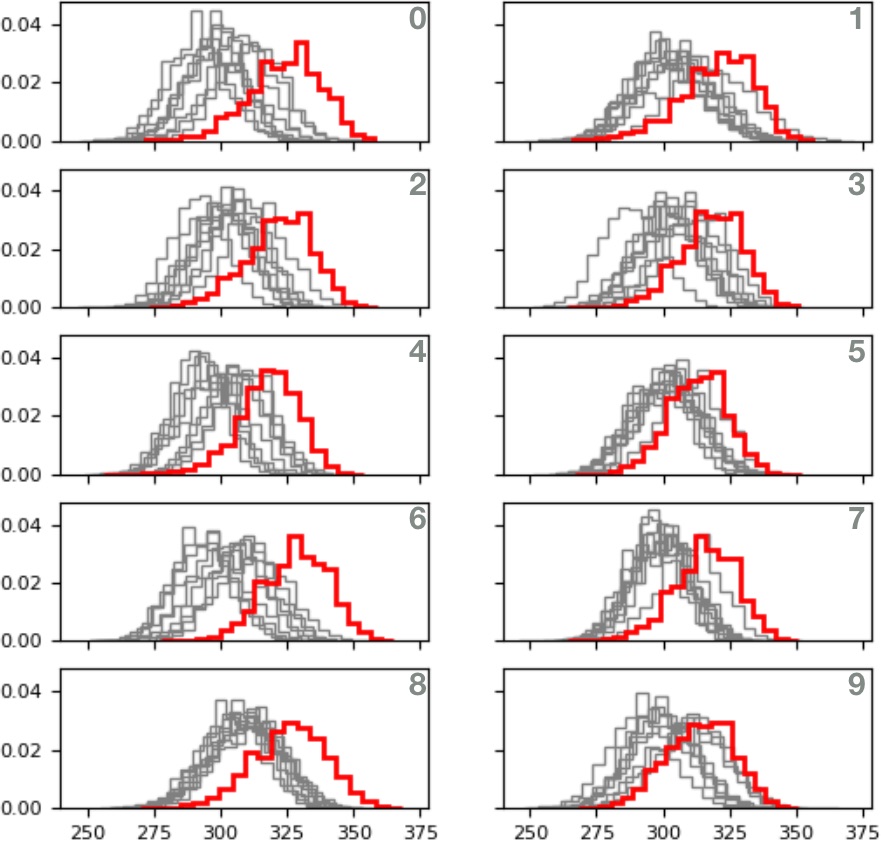}
\caption{The histograms of $Z$ for all ten digits. For each digit, randomly drawn testing image, instead of the same one, is used in each test. From digit $0$ to $9$, classification accuracy is approximately $70\%$, $41\%$, $56\%$, $42\%$, $53\%$, $33\%$, $77\%$, $51\%$, $57\%$ and $32\%$. Generally, better $Z$-distribution separation of positive and nagative testing results in higher classification accuracy. }
\label{fig:mesh_Z_dist}
\end{figure}

\begin{figure}[!ht]
\centering
\includegraphics[width=\linewidth]{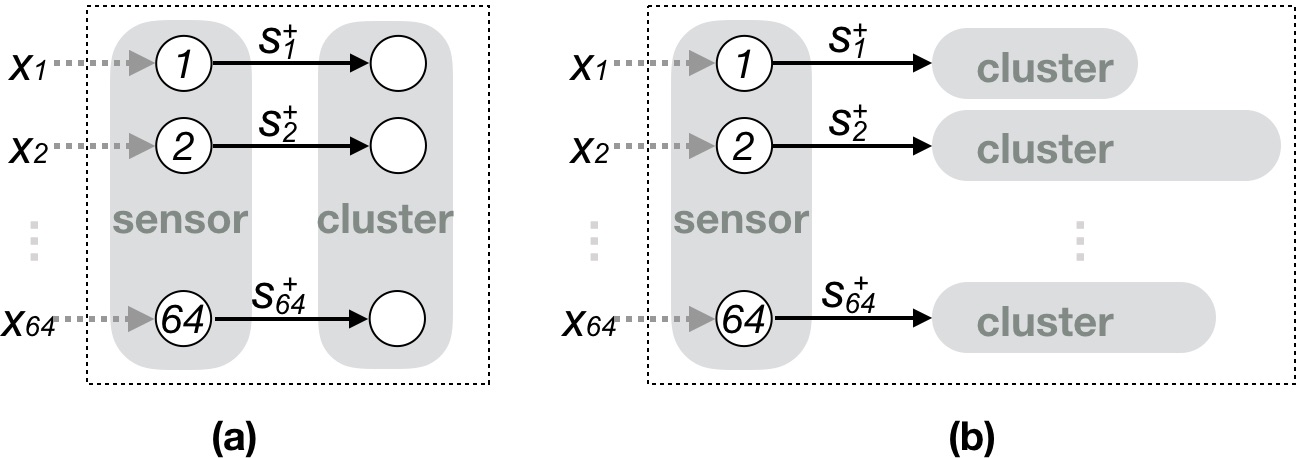}
\caption{These two neural networks inherit the sensor-cluster structure of FIG \ref{fig:brain1}.
\textbf{(a)} Each sensor neuron connects to one single cluster neuron such that each pixel stimulus $x_i$ only affects one single connection. Then $s_i^+{=}\theta_i(\check{x_i})$. By Eq. (\ref{Z_ev}) and Eq. (\ref{Z_var}), we have $\E[Z_{\check{X}X}]{=}\sum^{64}x_i\theta_i(\check{x_i})$ and $\Var(Z_{\check{X}X}){\approx}\sum^{64}x_i\theta_i(\check{x_i})[1-x_i\theta_i(\check{x_i})]$.
\textbf{(b)} Each sensor neuron connects to its own dedicated cluster of many neurons and synaptic connections, and the clusters are of different sizes.
    In that case, in a test each $x_i$ causes ${\omega}_i$ (instead of just one) synaptic connections to be propagated with probability $x_is_i^+$ or none with probability $1{-}x_is_i^+$.
    When each ${\omega}_i$ is a nonrandom variable, we have $\E[Z_{\check{X}X}]{=}\sum^{64}x_i\theta_i(\check{x_i}){\omega}_i$ and $\Var(Z_{\check{X}X}){\approx}\sum^{64}x_i\theta_i(\check{x_i})[1-x_i\theta_i(\check{x_i})]{\omega}_i^2$. }
\label{fig:brain2}
\end{figure}

We experiment with this classifier to classify handwritten digit images\footnote{The dataset of $1797$ handwritten digit images can be obtained with Python code \emph{"from sklearn import datasets; datasets.load\_digits()"}.}.
Ten identical neural networks (hence $g{=}10$) of FIG \ref{fig:brain1}, each designated for a digit from $0$ to $9$, are trained to their fixed points by their training images in FIG \ref{fig:average_images} as stimulus, and then testing images, also as stimulus, are classified into the digit whose designated neural network gets the biggest $Z$ value.
We run many tests to evaluate classification accuracy, and collect $Z$ values to approximate r.v. $Z$'s distribution.
With all synaptic connections equipped with $\lambda_{L}$ in FIG \ref{fig:best_lambda}, the classifier has accuracy ${\sim}44\%$, and ${\sim}51\%$ with $\lambda_{T}$.
Note that, equipped with $\lambda_{L}$ or $\lambda_{T}$, the neural network of FIG \ref{fig:brain1} will have one-to-one $\Theta_{L}$ or $\Theta_{T}$ according to last section.
FIG \ref{fig:Z_dist_of_6} and FIG \ref{fig:mesh_Z_dist} show that, in positive testing (e.g. digit-6 image is tested in neural network trained by digit-6 images), $Z$'s expected value (sample mean) could be considerably bigger than that in negative testing (e.g. digit-6 image is tested in neural network trained by digit-1 images), so as to discriminate digit-6 images from the others.
Given the same testing image classification target can be different test by test since the ten $Z$ outcomes are randomized.
To improve classification accuracy, we shall distance the distribution of positive testing $Z$ as far as possible from those of negative testing $Z$.
We present another two special neural networks in FIG \ref{fig:brain2} to demonstrate how our classifier utilizes memory to classify images and how to improve its accuracy in the neurobiological way.

\begin{figure}[!ht]
\centering
\includegraphics[width=\linewidth]{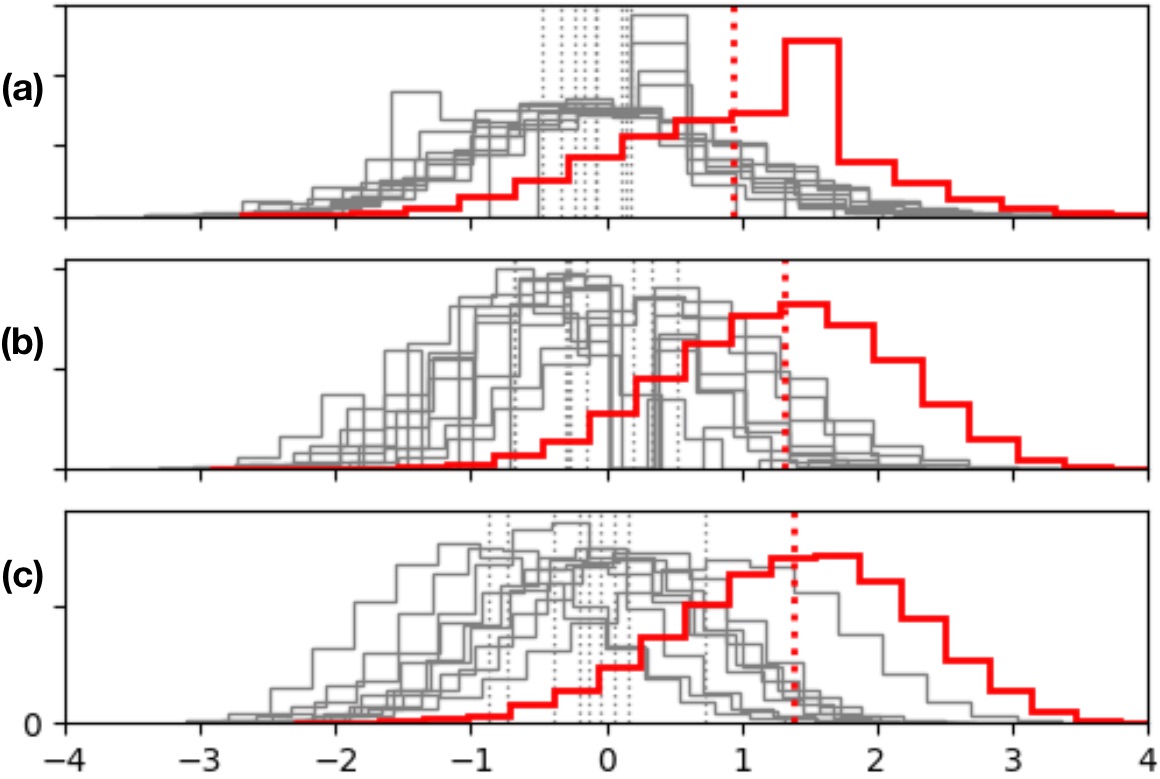}
\caption{The $Z$ histogram of testing digit-6 images for three different classifier settings.
For each classifier setting, values of $Z_{66}$ and $Z_{k6,k\neq6}$ are transformed to z-scores (i.e. the number of standard deviations from the mean a data point is) with respect to the distribution of all $Z_{k6,k\neq6}$'s values combined.
The distance between the distribution of $Z_{66}$ and $Z_{k6,k\neq6}$ is approximately evaluated by $\E[Z_{66}]$ and standard deviation $\sigma(Z_{66})$.
\textbf{(a)} Classifier with neural networks of FIG \ref{fig:brain2}(a) and $\lambda_L$. $\E[Z_{66}]{\approx}0.93$ and $\sigma(Z_{66}){\approx}0.98$.
\textbf{(b)} Classifier with neural networks of FIG \ref{fig:brain2}(a) and $\lambda_T$. $\E[Z_{66}]{\approx}1.33$ and $\sigma(Z_{66}){\approx}0.9$.
\textbf{(c)} Classifier with neural networks of FIG \ref{fig:brain2}(b), $\lambda_L$ and $\omega_i(\check{x_i}){=}100\check{x_i}^3$. $\E[Z_{66}]{\approx}1.39$ and $\sigma(Z_{66}){\approx}0.87$.}
\label{fig:Z_ev_variance}
\end{figure}

When the classifier adopts ten neural networks of FIG \ref{fig:brain2}(a) and equips all connections with $\lambda_{L}$ in FIG \ref{fig:best_lambda}, classification accuracy is ${\sim}31\%$ and $Z$'s distribution for testing digit-6 images is shown in FIG \ref{fig:Z_ev_variance}(a). We already know that $\lambda_{L}$ makes $\theta_L(x){\approx}x$. Then for one test we have

\begin{equation}
\E[Z_{\check{X}X}]=\sum^{64}x_i\theta_i(\check{x_i})=\sum^{64}x_i\theta_L(\check{x_i})\approx\sum^{64}x_i\check{x_i}=\check{X}^{\intercal}X.
\label{dot_product}
\end{equation}

Here $\check{X}^{\intercal}X$ is the dot product of training vector $\check{X}{\in}[0,1]^{64}$ and testing vector $X{\in}[0,1]^{64}$.
Generally, the dot product of two vectors, a scalar value, is essentially a measure of similarity between the vectors.
The bigger $\E[Z_{\check{X}X}]$ is, the more intensely neural network with memory of training $\check{X}$ responses to testing $X$, and the more similar $\check{X}$ and $X$ are to each other.
Therefore, Eq. (\ref{dot_product}) simply links otherwise unrelated neural response intensity and stimulus similarity.
By comparing ten $\E[Z_{\check{X}X}]$ values, we can tell which $\check{X}$ is the most similar to $X$ and hence which digit is classification target.
Only, $Z_{\check{X}X}$ value from test actually deviates around the true $\E[Z_{\check{X}X}]$ randomly, which makes it a useable and yet unreliable classification criteria.

When the classifier equips all connections with threshold-like $\lambda_{T}$ in FIG \ref{fig:best_lambda}, classification accuracy raises to ${\sim}44\%$. By comparing FIG \ref{fig:Z_ev_variance}(b) with FIG \ref{fig:Z_ev_variance}(a), the distance between $Z_{66}$'s distribution and the other nine $Z_{k6,k\neq6}$'s distribution is bigger with threshold-like $\lambda_{T}$ than with linear-like $\lambda_{L}$. This accuracy improvement can be explained conveniently with a true threshold function (or step function)

\begin{equation*}
\theta_{step}(x)=
\begin{cases}
0, & 0 \leq x < x_{step} \\
1, & x_{step} \leq x \leq 1
\end{cases}.
\end{equation*}

Of the sum terms in $\sum^{64}x_i\check{x_i}$ of Eq. (\ref{dot_product}), $\theta_{step}$ basically diminishes small $\check{x}_i{\in}[0,x_{step})$ to $0$ and enhances big $\check{x}_i{\in}[x_{step},1]$ to $1$, such that most probably $\E[Z_{66}]$ would increase by having $\check{x}_i{=}1$ in the sum terms with big $x_i$ while the other nine $\E[Z_{k6,k\neq6}]$ would decrease by having $\check{x}_i{=}0$ in the sum terms with big $x_i$, so as to preferably increase $\E[Z_{66}]{-}\E[Z_{k6,k\neq6}]$.
And likewise $\Var(Z)$ would most probably decrease. As a result, $\theta_{step}$ increases the distance between the distribution of $Z_{66}$ and $Z_{k6,k\neq6}$ and thus better separates them.

FIG \ref{fig:brain2}(b) provides another type of neural network to improve classification accuracy without adopting threshold-like $\lambda$ function for all synaptic connections.
Let the linear-like $\lambda_L$ be equipped back and take ${\omega}_i{=}100\check{x_i}^3$ simply for example. With this setting our classifier has accuracy ${\sim}47\%$.
Here we have $\E[Z_{\check{X}X}]{\approx}\sum^{64}x_i(100\check{x_i}^4)$ where $100\check{x_i}^4$, like $\theta_{step}$, actually transforms $\check{x_i}{\in}[0,\sqrt[4]{0.1})$ (here $\sqrt[4]{0.1}{\approx}0.56$) to within $[0,10)$ and transforms $\check{x_i}{\in}[\sqrt[4]{0.1},1]$ to across $[10, 100]$ — again, the strong training pixel-stimulus are greatly weighted while the weak ones are relatively suppressed.
As shown in FIG \ref{fig:Z_ev_variance}(c) the distance between the distribution of $Z_{66}$ and $Z_{k6,k\neq6}$ is increased compared to FIG \ref{fig:Z_ev_variance}(a).
Here our neurobiological interpretation regarding ${\omega}_i{=}100\check{x_i}^3$ is, the training stimulus affects not only synaptic strength, but also the growth of neuron cluster in the replication of neuron cells and in the formation of new synaptic connections. Again this claim needs to be supported by evidences.

TABLE \ref{tab:table1} summarizes the performance of our classifier with different types of neural networks and target strength functions. The four typical $\lambda$ functions in FIG \ref{fig:tendency_simulation} are also evaluated to demonstrate how these somewhat "pathological" target strength functions affect classification.

\begin{table}
\caption{\label{tab:table1}Classification accuracy on different classifier settings. In each setting, neural network can be the one illustrated in FIG \ref{fig:brain1}, FIG \ref{fig:brain2}(a) or \ref{fig:brain2}(b), and target strength function can be one of those in FIG \ref{fig:tendency_simulation} and FIG \ref{fig:best_lambda}. The accuracy listed is the average of many outcomes taken from the same trained classifier, and thus could fluctuate slightly from one training to another. }
\begin{ruledtabular}
\begin{tabular}{llll}
$\lambda$ or $\theta$ functions &FIG \ref{fig:brain1}&FIG \ref{fig:brain2}(a)&FIG \ref{fig:brain2}(b)\\
\hline
$\lambda_L(y){=}0.99\sqrt{y}{+}0.01$ & $44\%$ & $31\%$ & $47\%$\\
$\lambda_T(y){=}\frac{2}{1{+}e^{{-}4.4(y{+}0.01)}}{-}1$ & $51\%$ & $44\%$ & $51\%$ \\
$\theta_{step}$ & - & $48\%$\footnote{$x_{step}$ is set to $0.6$.} & $60\%$ \footnote{$x_{step}$ is set to $0.2$.} \\
\hline
$\lambda(y){=}0.9y{+}0.05$ & $14\%$ & $16\%$ & $19\%$\\
$\lambda(y){=}0.5sin(4{\pi}y){+}0.5$ & $5\%$\footnote{Accuracy under $10\%$ is actually worse than wild guessing. } & $6\%$ & $2\%$\\
$\lambda(y){=}{-}y{+}1$ & $4\%$ & $5\%$ & $1\%$\footnote{If classification criteria is changed to "which neural network gets the fewest synaptic connections propagated", the accuracy will be ${\sim}40\%$. }\\
Discontinuous $\lambda$ & $23\%$ & $20\%$ & $28\%$\\
\end{tabular}
\end{ruledtabular}
\end{table}

By Eq. (\ref{dot_product}), the classification of handwritten digit images could be simplified to a task of restricted linear classification \cite{Element}: given ten classes each with its discriminative function $\delta_i(X){=}\check{X}_i^{\intercal}X$ where $\check{X},X{\in}[0,1]^{64}$, image $X$ is classified to the class with the largest $\delta_i$ value.
Our neural classifier simply takes over the computation of vectors' dot product $\check{X}_i^{\intercal}X$ and adds randomness to the ten results.
To parameterize the ten $\delta_i$ with their $\check{X}_i$, the "supervisors" could train the neural networks in classifier with the images they deem best — "average images" in our case or digits learning cards in teachers' case.
Our neural classifier is rather unreliable and primitive compared to ANN which is also capable of linear classification.
On one hand, given the same image ANN always outputs the same prediction result.
On the other hand, ANN is not only a classifier but also more importantly a "learner", which learns from all kinds of handwritten digits to find the optimal $\check{X}_i$ for the ten $\delta_i$; ANN with optimal $\check{X}_i$ is more tolerant with poor handwriting, and thus has less misclassification and better prediction accuracy.
Only, ANN's learning optimal $\check{X}_i$, an optimization process of many iterations, requires massive computational power to carry out, which is unlikely to be provided by the real-life nervous system — there is no evidence that an individual neuron can even conduct basic arithmetic operations.
Despite of its weakness, our neural classifier has merit in its biological nature:
it reduces the computation of vectors' dot product to simple counting of synaptic connections propagated;
its training and testing could be purely neurobiological development and activities where no arithmetic operation is involved;
its classification criteria, i.e. "deciding" or "feeling" which neural (sub)network has the most connections propagated, could be an intrinsic capability of intelligent agents.
This classifier might project new insights on the neural reality, hopefully.

\section{Conclusion}
\label{section_conclusion}

This paper proposes a mathematical theory to explain how memory forms and works.
It all begins with synaptic plasticity.
We find out that, synaptic plasticity is more than impulses affecting synapses; it actually plays as a force that can drive neural network eventually to a long-lasting state.
We also find out that, under certain conditions there would be a one-to-one mapping between the neural state and the external stimulus that neural network is exposed to.
With the mapping, given stimulus we know exactly what neural state will be; given neural state we know precisely what stimulus has been.
The mapping is essentially a link between past event and neural present; between the short-lived and the enduring.
In that sense, the mapping itself is memory, or the mapping casts memory in neural network.
Next, we study how memory affects neural network's response to stimulus.
We find out that, the neural network with memory of stimulus can response to similar stimulus more intensely than to the stimulus of less similarity, if response intensity is evaluated by the number of synaptic connections propagated by impulses.
That is to say, a neural network with memory is able to classify stimulus.
To verify this ability, we experiment with the classifier consisting of ten neural networks, and they turn out to have considerable accuracy in classifying the handwritten digit images.
The classifier proves that neurons could collectively provide fully biological computation for classification.

Our reasoning takes root in the mathematical treatment of synaptic plasticity as target strength function $\lambda$ from impulse frequency to synaptic strength.
We put hypothetical constraints on this $\lambda$ function to ensure that the ideal one-to-one mapping exists.
Although these constraints are necessary to keep our theory mathematically sound, they raise concerns.
Firstly, they could be overly restrictive.
Take continuity constraint for example.
Even the discontinuous function of FIG \ref{fig:tendency_simulation}(d), whose nonexistent $\theta$ function would map certain stimulus to any point within a "fixed interval" instead of a specific fixed point as shown in FIG \ref{fig:theta_discont_pf}, can be a useable $\lambda$ in our classifier according to TABLE \ref{tab:table1}.
In this case, fixed point per se doesn't have to exist, and mere tendency to seek out for it could serve the purpose.
Secondly, as discussed in Section \ref{section_connection} those $\lambda$ constraints have yet to be supported by neurobiological evidences.
Above all, the evidence that reveals true $\lambda$ is vital to clarify the uncertainty.

\section{Acknowledgement}
We thank the anonymous reviewers for their comments that improved the manuscript.

\bibliographystyle{unsrt}
\bibliography{refs}
\end{document}